\documentclass[12pt,preprint]{aastex}

\newcommand{\HI}{\ion{H}{1}\ }
\newcommand{\hi}{\ion{H}{1}\ }
\newcommand{\halpha}{H$\alpha$}
\newcommand{\vlsr}{\ensuremath{v_{\mathrm{LSR}}}}
\newcommand{\kms}{\ensuremath{\mathrm{km\ s}^{-1}}\ }
\newcommand{\kmss}{\ensuremath{\mathrm{km\ s}^{-1}}}
\newcommand{\Trms}{\ensuremath{\Delta\mathrm{T}_{\mathrm{rms}}}}
\newcommand{\co}{\ensuremath{^{12}\mathrm{CO}}}
\newcommand{\kpc}{kpc}
\newcommand{\paper}{\emph{Letter}}

\newcommand{\hh}{\ensuremath{\mathrm{H}_2}}

\shorttitle{Tycho: possible interaction with MC}
\shortauthors{Lee, Koo, \& Tatemastu}

\begin{document}

\title{The environment of Tycho: possible interaction with molecular cloud}

\author{Jae-Joon Lee and Bon-Chul Koo}
\affil{Astronomy program, SEES, Seoul National University, Seoul 151-742, Korea }
\email{jjlee@astro.snu.ac.kr, koo@astrohi.snu.ac.kr}

\and

\author{Keni'chi Tatemastu}
\affil{National Astronomy Observatory of Japan, 2-21-1 Osawa, Mitaka, Tokyo, 181-8588, Japan}
\email{k.tatematsu@nao.ac.jp}

\begin{abstract}
The Tycho supernova remnant (SNR), as one of the few historical SNRs, has been
widely studied in various wavebands. Observations
show evidence that Tycho is
expanding in a medium with density gradient and
possibly  interacting with dense ambient medium 
toward the northeast direction. From the FCRAO
CO survey of the outer Galaxy, we have identified a patch
of molecular clouds 
in this area and
have conducted a follow-up observation with Nobeyama 45m radio telescope.
The high-resolution (16\arcsec) Nobeyama data shows 
that 
a large molecular cloud surrounds 
the SNR along the northeastern boundary.
We suggest that the Tycho SNR and the molecular cloud
are located in the Perseus arm and that
the dense medium interacting with the SNR is possibly 
the molecular cloud.
We also discuss the possible connection between the molecular 
cloud and the Balmer-dominated optical filaments, and suggest
that the preshock gas may be accelerated within the cosmic ray 
and/or fast neutral precursor.

\end{abstract}


\section{INTRODUCTION}

The Tycho supernova remnant (SNR)
 is known as a remnant of a Type Ia supernova that occurred
in the year 1572 \citep{1945ApJ...102..309B}. 
As one of the few historical SNRs, the remnant 
has been widely studied in various wavebands.
In radio continuum, it shows a shell morphology
with clearly defined outer boundary and clumpy
inner structure \citep[e.g.,][]{1991AJ....101.2151D
}.
Recent 
high resolution X-ray images
 also exhibit similar features 
\citep{2001A&A...365L.218D,2002ApJ...581.1101H}.
\cite{1980ApJ...235..186C} 
derived a shock velocity by modelling the optical spectrum
from a fast nonradiative shock and compared it
with an optical proper motion 
\citep{1978ApJ...224..851K}
, giving an estimated distance of $2.3 \pm 0.5$ \kpc.
Other independent estimates give similar distances
\citep{1986MNRAS.219..427A,1988MNRAS.230..331S}.
It is, however, noted that there has been
controversy in interpreting the \hi absorption data, 
giving a different distance \citep[e.g.,][]{1995A&A...299..193S}.
This will be discussed in \S\ \ref{distance}.

Observations show evidence
that Tycho is interacting
with ambient dense clouds toward its northeast (NE) direction. 
\citet{1997ApJ...491..816R} has studied its expansion
using a 10-year-lap VLA data
and found that the NE area, 
which is farthest from the geometrical center of the SNR, is currently
moving slowest, suggesting that the remnant is expanding into a 
higher density medium.
\citet{1999AJ....117.1827R} 
has conducted a \HI absorption study
towards the remnant 
and
suggested that Tycho is currently interacting with
a dense \HI concentration in the NE
($\vlsr = -51.5$\ \kms) which locally slows down the expansion
of the shock front.
There are other observations that also suggest its interaction
with inhomogeneous environment 
\citep{
1998A&A...334.1060V,
2000ApJ...529..453K,
2001A&A...373..281D}.

From the FCRAO $^{12}$CO survey of the outer Galaxy 
\citep{1998ApJS..115..241H},
 we have identified
molecular clouds 
which could be associated with Tycho (Figure \ref{fig:tycho_fcrao}).
They are arc-shaped and the inner boundary 
roughly traces the NE rim of Tycho.
The morphology of the molecular clouds 
suggests physical
association with Tycho. 
We have conducted 
high-resolution $^{12}$CO J=$1\!-\!0$ line observations
and present the results in this \paper.

\section{OBSERVATION AND RESULT}
The observation was carried out for a total of
20 hours during 2003 January 11--13 using the Nobeyama
45m radio telescope. The BEARS focal-plane
receiver system was used to cover
$12\arcmin\times 12\arcmin$\ area centered at Tycho whose diameter 
is $\sim 8\arcmin$.
HPBW of the 45m telescope at $^{12}$CO (J=1-0) is about $16\arcsec$ and
we have mapped the region with a 8\arcsec\ grid spacing.
Pointing accuracy was quite good during the observing run 
and we estimate it to be  better than 5\arcsec.
Velocity resolution of $0.08$ \kms was
achieved using 1024 channel autocorrelator as a backend.
The data were calibrated using the standard procedure with
\emph{NewStar} software developed by NAOJ. 
Exposure time is not uniform, leading to varying noises. 
The NE portion of the image
has the most significant S/N ($\Trms \sim 0.1\ \mathrm{K}$)
while the inner part
of the remnant has a poorer S/N ratio ($\Trms = 0.3\ \mathrm{K}$).

In Figure \ref{fig:tycho_channelmap_nobeyama1}, 
we present channel maps in gray scale
together with the radio continuum showing the boundary of Tycho.
There is virtually no \co\ emission
at $\vlsr > -55\ \kms$\ (except from the gas in the local arm)
and most of the emission is between
$-67\ \kms < \vlsr < -60\ \kms$
with some faint emission localized in the northern part of the remnant 
at $\vlsr = -58 \sim -55\ \kmss$.
The emission is generally from regions surrounding the remnant.
In particular, at velocities between $-63.5$ to $-61.5\ \kmss$,
the emitting area appears to be in contact with the remnant
along its NE boundary. This is more clearly seen in 
Figure\ \ref{fig:tycho_v63_nobeyama}, the integrated 
intensity map over $\vlsr = -63\sim -60\ \kmss$.
In particular, at the  location of optical knot g where 
the radio continuum boundary is deformed and the 
radio expansion 
is slowest, molecular clumps are touching the radio boundary.

\section{DISCUSSION}
The ambient environment of Tycho is not homogeneous 
as clearly revealed by the slower expansion of the NE rim.
The morphology of the molecular cloud suggests
their physical interaction. 
We discuss their physical association in the following sections.

\subsection{Distance to the Tycho SNR and the Molecular Cloud
\label{distance}
}
The distance to Tycho has been estimated by several authors using 
independent methods 
\citep{1978ApJ...224..851K, 
       1986MNRAS.219..427A,
       1988MNRAS.230..331S}.
Most estimates agree on 
a distance of $\sim 2.3$ kpc except for the estimate based on \hi 
absorption observations. Applying the \hi absorption technique to 
Tycho is not straightforward due to the non-circular motion 
of the gas in the Perseus arm. The peculiar velocity field of the 
Perseus arm is well known, e.g., radial velocities of \ion{H}{2} 
regions and OB stars in the Perseus arm are lower than those derived
from photometric/spectroscopic distances 
\citep[see][and reference therein]{1995A&A...299..193S}
.
The widely accepted model is the
``two armed spiral shock'' (TASS)
model proposed by \cite{1972ApJ...173..259R}
. In this model, the Galactic spiral pattern is visualized 
by a large scale shock residing in the Galactic density wave.   
Multi-value velocity-distance relation is naturally expected in this 
scenario, and,  toward Tycho ($\ell=120\arcdeg$), 
the radial velocity is expected to
drop abruptly from $-20\ \kms$ to $-50~(\pm 8)$ \kms\
at the shock front at 2.2 kpc 
and slowly increase to $-40~\kms$\ at 2.8 kpc, after which it slowly 
decreases \citep[see Figure 14 of][]{1995A&A...299..193S}.

The velocity of the most negative absorption feature towards 
Tycho is around $-50\ \kmss$ with weak absorption at $-59\ \kmss$.
\cite{1986MNRAS.219..427A} 
attributed the absorption to the gas behind the spiral shock and 
placed Tycho just behind the large scale spiral shock, at a distance of 
$2.2^{+1.5}_{-0.5}$\ kpc. They noted that there is little absorption 
between $-45$ and $-38$~\kms\ which was considered as evidence 
for Tycho not being far beyond the spiral shock front. On the other hand,
\cite{1995A&A...299..193S} 
has argued that it should be placed at a distance of $4.6\pm 0.5$ 
kpc where almost circular rotation is applicable.
One of their arguments was that the velocity of the 
most negative absorption feature ($-59\ \kmss$) 
is  too low compared
to the model-predicted shock velocity ($-50$ \kmss). 
The velocity structure in the Perseus arm, however, is rather highly 
dependent
on the Galactic latitude. This trend can be clearly seen
from the recent FCRAO CO survey of the outer Galaxy 
\citep{1998ApJS..115..241H}
\ and DRAO low resolution \hi\ survey 
\citep{2000AJ....120.2471H}
. 
The peak velocity of \co\ associated with Perseus arm
in the region of $l \sim 120\arcdeg$\ is
around $\sim -50\ \kms$ at $b \sim 0\arcdeg$, but  
systematically decreases to 
$\sim -60\ \kms$\ at $b \sim 1\arcdeg$\ and further decreases
for higher latitudes.
If these \co\ molecules are associated with the high density
regime behind the spiral shock front, as is likely to,
the shock velocity itself
should have a similar trend, although 
it is not explicitly 
accounted for in the TASS model.
\hi data can be described in a similar way.
Therefore, if we account the $b$-dependent behavior of the 
velocity field, the absorption velocity could be as low as $-60\ \kms$
even if Tycho is just behind the shock front, and we consider that the 
\hi absorption experiments are consistent with Tycho being at the near 
side of the Perseus arm.  

The velocity of the CO molecular cloud in the NE area of 
Tycho is $\sim -62$ km/s, which is slightly lower than the minimum 
velocity of the \hi absorption feature. The cloud, and other 
clouds in 
Figure \ref{fig:tycho_channelmap_nobeyama1}, however, seem to be part of a 
large-scale distribution of molecular gas presumably representing 
the high-density ridge behind the spiral shock. The CO emission 
between $\ell=119\arcdeg$ and $121\arcdeg$ is generally confined within 
a velocity interval of $-66$ and $-60\ \kmss$. 
This is much narrower than that  of 
the \hi gas ($-70\sim-45 \kmss$), whose
peak velocity is also blue-shifted from the peak of \co.
These different characteristics of \co\ and \hi 
can be explained within the
frame of TASS model where 
\co\ clouds mostly correspond to the high-density region
just behind the spiral shock.
We have compared the $l-v$ distribution of \co\ around 
$b \sim 0\arcdeg$ with the theoretical prediction (shock ridge)
and found them to be well matched.
Hence, it is likely that the distances to the Tycho SNR and to the 
molecular clouds surrounding it
are comparable.

\subsection{Non-radiative \halpha\ filaments}

Optical filaments have been found in Tycho
\citep{1991ApJ...375..652S}
. The filaments are located along the NE boundary 
of the remnant, where interaction with a molecular cloud is likely
(Figure \ref{fig:tycho_v63_nobeyama}). 
The emission is predominantly hydrogen Balmer emission and
is believed to originate from a
collisionless nonradiative shock, characterized by
broad and narrow emission components whose velocity distribution
represent that of postshock and preshock gas, respectively
\citep{
1978ApJ...225L..27C
,1980ApJ...235..186C
}. 
The expansion velocities of these filaments, 
which are compatible with radio expansion rate at
this region 
\citep
[$\sim 0.15 \arcsec/\mathrm{yr}$,][]{1997ApJ...491..816R}
, is relatively small ($\sim 30\ \%$) compared to 
the other parts of the remnant.
The implied large preshock density suggests its association
with the molecular cloud.
\cite{2000ApJ...535..266G}
\ obtained high-resolution echelle 
spectrum of knot g (Figure \ref{fig:tycho_v63_nobeyama}) and
their observed central velocity
is $\vlsr = - 53.9 \pm 1.3\ \kmss$.
For comparison, most of the \co\ emission around Tycho is between 
$-67$ and $-60\ \kmss$, and the 
\co\ line profile near knot g, 
fitted with Gaussian,
gives a line center of $-62.50 \pm 0.01\ \kmss$\ and a FWHM of 
$2.27 \pm 0.03\ \kmss$. Therefore, there is a non-negligible 
velocity difference of $8.6 \pm 1.3\ \kmss$ between the 
H alpha filament and the \co\ cloud. Even if we consider the FWHM 
of \co\ as a possible error, the difference is greater than 
$6\ \kms$ which is significant.
The velocity of the narrow \halpha\ component of a non-radiative 
shock should represent that of the gas entering the shock. 
This implies two possibilities. One is that the gases responsible for
\co\ and \halpha\ have initially different velocities.
This could be due to a turbulent motion behind the spiral shock or 
some kinematical effects associated with the evolution of the progenitor 
star or the SN explosion itself. But these seem rather unlikely because
there is virtually no \co\ emission above $-60\ \kmss$\ and 
the actual velocity difference considering a projection could be much
larger (see next section). 
The remaining possibility is that the gas experiences an acceleration
before the excitation by the shock. 
We further discuss about it in the following.

A handful of remnants show non-radiative shock emission like Tycho
\citep{
1991ApJ...375..652S
,1994ApJ...420..286S
,2001ApJ...547..995G
,2003A&A...407..249S
}.
The width of the narrow line has been a puzzle since it is unusually
large for neutral hydrogen and does not vary much from 
remnant to remnant.
For most of the remnants,
it ranges between 30 and 50 $\mathrm{km\ s}^{-1}$, while SN 1006, which 
is the fastest ($\sim 3000\ \kmss$), has a slightly 
smaller value of about 20 \kmss. 
Various physical mechanisms have been proposed to explain this line width
\citep{1994ApJ...420..286S}
. 
Although the CR precursor and/or fast neutral precursor
scenario has been regarded as the most likely case
\citep{1994ApJ...420..286S
,2000ApJ...535..266G}
,\ firm evidence is yet to be found.
Some of the proposed mechanisms, which include a interaction with
heavy particles (like protons), also predict a 
systematic acceleration of the gas before their excitation.
We propose that the velocity difference 
between \halpha\ and \co\ can be explained by this
gas acceleration.
Observed peak velocity of the broad \halpha\ component is red-shifted 
by $132\pm 35\ \kmss$ from 
that of the narrow one, indicating that the shock is not completely 
edge on \citep{2001ApJ...547..995G}
.
With a shock velocity $v_s = 2000\ \kmss$ (fluid velocity $\sim\frac34v_s$%
),
viewing angle is estimated to be about 
$\theta \sim 5 \arcdeg$.
If the gas is accelerated,
the \halpha\ narrow
component will be red-shifted from the unperturbed medium,
which is qualitatively consistent with 
our interpretation
where \co\ typifies the velocity of the unperturbed medium.
The observed velocity difference of $8.6 \pm 1.3\ \kms$ in \vlsr\ 
then implies that the actual acceleration is about
$ \Delta v = 98\pm 30\ ( {v_s}\,/\,{2000\ \kmss})\ \kmss$.
One possibility to explain the velocity difference is a
momentum transfer by collisions with fast protons 
after crossing the shock.
But, as \cite{1994ApJ...420..286S}
\ has showed, a lifetime of the hydrogen atom is very short
and only a moderate amount of 
acceleration ($\lesssim 10\ \kmss$) is possible.
We consider two preshock mechanisms which might account 
for this acceleration.

Broadening of the narrow component is well explained by a CR
precursor \citep{1994ApJ...420..286S}
\ and acceleration of the preshock gas is
also theoretically well expected
\citep[see][]{
1987PhR...154....1B
}.
For a CR dominated shock, it is suggested that
the preshock gas can be accelerated to the significant
fraction of the shock velocity.
\cite{1988ApJ...333..198B}
\ has applied this shock model
to the Balmer-dominated filaments in Cygnus loop.
According to their model, the preshock gas is accelerated to 
velocities close to 
the shock velocity.
For the case of Tycho,
the smaller amount of acceleration, compared to the shock velocity,
indicates that the cosmic ray contribution is not dominant.
This is consistent with recent X-ray results 
\citep{2002ApJ...581.1101H} where no significant temperature
drop due to a possible cosmic ray acceleration is observed.
The radio spectral index ($S_{\nu} \propto \nu^{-\alpha}$) 
of Tycho is known to be steeper 
($\alpha \sim 0.6$ at $\nu \ge 1\  \mathrm{GHz} $) than what is
expected from a test particle theory ($0.5$), and it is suggested that
Tycho may have somewhat strongly modified shock 
\citep{1992ApJ...399L..75R,2002A&A...396..649V}.
But, the spectral index varies over the remnant 
\citep{2000ApJ...529..453K}  and knot g region 
(filament I in \citeauthor{2000ApJ...529..453K})
has the flattest spectral index of $0.44 \pm 0.02$,
which might be another indication for the interaction
with molecular clouds 
\citep{
2001ApJ...552..175K}.
Although Tycho may have strongly modified shock 
in a large scale, it does not seem to apply to areas interacting with
molecular clouds.

There has been little
theoretical study on the fast neutral precursor
compared to the cosmic ray precursor.
\cite{1996MNRAS.280..103L} calculated 
the \halpha\ profile 
from a $225\ \kms$ non-radiative shock
with a fast neutral precursor, but due to their narrow
parameter range, direct comparison with
our observation is not available and there is no
indication of the velocity shift of the narrow component in their
line profiles. Calculations for higher shock velocity and lower
pre-ionization fraction is necessary.

\section{CONCLUSION}

The study of Balmer-dominated filaments in SNRs has
given us much information
about the shock structure. 
The velocity of an unperturbed preshock medium is an invaluable
observational ingredient for fully understanding underlying 
shock  structure.
In Tycho, a velocity shift between an ambient molecular cloud,
which is believed to be the unperturbed preshock gas, and
the \halpha\ filaments has been observed. 
Although the preshock acceleration was searched for previously,
no evidence has obtained before our observation
\citep[e.g.][]{1994PASP..106..780E}. Our observation serves as the first 
direct evidence for the acceleration of preshock gas
by a cosmic ray (or fast neutral) precursor. 
We, however, note that our conclusion is based on the single
velocity measurement and more observations are needed
to test it.

There has been not much study on the structure
of collisionless shocks propagating
into  molecular clouds. 
But most of the volume of a molecular cloud may be
occupied by low density ($\lesssim 10\ \mathrm{cm}^{-3}$) 
interclump medium,
and the structure would not much differ from 
that in atomic medium.
The molecules are supposed to be dissociated within a precursor
either by radiation from postshock region or hot electrons.
The dissociation rate of \hh\ is only slightly larger than
 ionization rate of the hydrogen atom
\citep{1998ApJ...496.1044Y,
2000A&AS..146..157L},
and complete dissociation of \hh\ may not happen
since they occupy more dense region.
The velocity of the transmitted shock into a molecular clump can be
sufficiently low to be radiative, 
but currently there is no sign of radiative shock in Tycho,
presumably because the shock is propagating into the low-density 
outskirts of the molecular cloud.
Future study should address these issues together
with the nature of cosmic ray acceleration
for a high-velocity ($\gtrsim 1000\ \kmss$) shock 
propagating into a molecular cloud.

\acknowledgments
We would like to thank Parviz Ghavamian for kindly providing
the \halpha\ data and useful comments.
We thank John Raymond, Hyesung Kang, Christopher McKee 
and Gloria Dubner for helpful
informations and comments. 
We also thank to the anonymous referee for valuable comments,
which helped us to improve this paper.
This work was supported by the 
Korea Science and Engineering Foundation
(ABRL 3345-20031017).
JJL has been supported in part by the BK 21 program.

\clearpage

\begin{figure}[t]
\caption{Integrated channel map of \co\ from 
FCRAO data ($ -68\le  v\le  -59\ \kmss$). 
White contour for  1420 MHz radio continuum. Black
boundary represents Nobeyama observed region.
\label{fig:tycho_fcrao}}
\end{figure}

\begin{figure}[t]

\caption{\co\ channel map of Nobeyama observation with 
$ \Delta v = 1\ \kmss$. Contour represents 1420 MHz radio continuum.
The last panel shows average spectrum of total observed region.
The peak brightness temperature of the
average spectrum is $0.83$ K.
\label{fig:tycho_channelmap_nobeyama1}
}
\end{figure}

\begin{figure}[t]

\caption{Pseudo color image of 
Integrated channel map ($ -63\le  v\le  -60\ \kmss$).
White contour for 1420 MHz radio continuum, 
red for \halpha\ \citep{2000ApJ...535..266G},
and blue for hard X-ray ($4$ keV $< h\nu< 6$ keV).
\label{fig:tycho_v63_nobeyama}}
\end{figure}

\clearpage 

\end{document}